\documentclass[12pt]{article}

\begin{document}

\title{\bf Kinematic Self-Similar Cylindrically Symmetric Solutions}

\author{M. Sharif \thanks{msharif@math.pu.edu.pk} and Sehar Aziz
\thanks{sehar$\_$aziz@yahoo.com}\\
Department of Mathematics, University of the Punjab,\\
Quaid-e-Azam Campus, Lahore-54590, Pakistan.}

\date{}

\maketitle

\begin{abstract}
This paper is devoted to find out cylindrically symmetric
kinematic self-similar perfect fluid and dust solutions. We study
the cylindrically symmetric solutions which admit kinematic
self-similar vectors of second, zeroth and infinite kinds, not
only for the tilted fluid case but also for the parallel and
orthogonal cases. It is found that the parallel case gives
contradiction both in perfect fluid and dust cases. The orthogonal
perfect fluid case yields a vacuum solution while the orthogonal
dust case gives contradiction. It is worth mentioning that the
tilted case provides solution both for the perfect as well as dust
cases.
\end{abstract}


\section{Introduction}

Due to the mathematical complexity of Einstein field equations
(EFEs)
\begin{eqnarray}
R_{ab}-\frac{1}{2}g_{ab}R=8\pi G T_{ab},
\end{eqnarray}
we are frequently forced to impose some symmetry on the concerned
system. Self-similarity is very helpful in simplifying the field
equations. In Newtonian gravity or General Relativity (GR), there
does not exist any characteristic scale. A set of field equations
remains invariant under a scale transformation for an appropriate
matter field. This indicates that there exist scale invariant
solutions to the EFEs. These solutions are known as self-similar
solutions. The special feature of self-similar solutions is that,
by a suitable coordinate transformations, the number of
independent variables can be reduced by one and hence reduces the
field equations. This variable is a dimensionless combination of
the independent variables, namely the space coordinates and the
time. In other words, self-similarity refers to an invariance
which simply allows the reduction of a system of partial
differential equations to ordinary differential equations.

In order to obtain realistic solutions of gravitational collapse
leading to star formation, self-similar solutions have been
investigated by many authors in Newtonian gravity [1]. However, in
GR, these solutions were first studied by Cahill and Taub [2].
They studied these solutions in the cosmological context and under
the assumption of spherically symmetric distribution of a
self-gravitating perfect fluid. In GR, self-similarity is defined
by the existence of a homothetic vector (HV) field. Such
similarity is called the first kind (or homothety or continuous
self similarity (CSS)). There exists a natural generalization of
homothety called kinematic self-similarity, which is defined by
the existence of a kinematic self-similar (KSS) vector field. The
basic condition characterizing a manifold vector field $\xi$ as a
self-similar generator is given by
\begin{eqnarray}
\pounds_\xi A=\lambda A,
\end{eqnarray}
where $\lambda$ is a constant, $A$ is independent physical field
and $\pounds_{\xi}$ denotes the Lie derivative along $\xi$. This
field can be scalar (e.g., $\mu$), vector (e.g., $u_{a}$) or
tensor (e.g., $g_{ab}$).  In GR, the gravitational field is
represented by the metric tensor $g_{ab}$, and an appropriate
definition of geometrical self-similarity is necessary.

The self-similar idea of Cahil and Taub corresponds to Newtonian
self-similarity of the homothetic class. Carter et al. [3,4]
defined the other kinds of self-similarity namely second, zeroth
and infinite kind. In the context of kinematic self-similarity,
homothety is considered as the first kind. Several authors [5-12]
have explored KSS perfect fluid solutions. The only barotropic
equation of state which is compatible with self-similarity of
first kind is
\begin{eqnarray}
p=k\rho.
\end{eqnarray}
Carr [5] has classified the self-similar perfect fluid solutions
of the first kind for the dust case ($k=0$). The case $0<k<1$ has
been studied by Carr et al. [6]. Coley [7] has shown that the FRW
solution is the only spherically symmetric homothetic perfect
fluid solution in the parallel case. McIntosh [8] has discussed
that a stiff fluid ($k=1$) is the only compatible perfect fluid
with the homothety in the orthogonal case. Benoit et al. [9] have
studied analytic spherically symmetric solutions of the EFEs
coupled with a perfect fluid and admitting a KSS vector of the
first, second and zeroth kinds.

Carr et al. [10] have considered the KSS associated with the
critical behavior observed in the gravitational collapse of
spherically symmetric perfect fluid with equation of state
$p=k\rho$. They showed for the first time the global nature of
these solutions and showed that it is sensitive to the value of
$\alpha$ (given in Eq.(15)). Carr et al. [11], further,
investigated solution space of self-similar spherically symmetric
perfect fluid models and physical aspects of these solutions. They
combine the state space description of the homothetic approach
with the use of the physically interesting quantities arising in
the co-moving approach. Coley and Goliath [12] have investigated
self-similar spherically symmetric cosmological models with a
perfect fluid and a scalar field with an exponential potential.

Gravitational collapse is one of the fundamental problems in GR.
Self-similar gravitational collapse and critical collapse provides
information about the collapse. The collapse generally has three
kinds of possible final states. First is simply the halt of the
processes in a self-sustained object or the description of a
matter field or gravitational field. The second is the formation
of black holes with outgoing gravitational radiation and matter,
while the third is the formation of naked singularities. Critical
collapse in the context of self-similarity gives the information
about the mass of black holes formed as a result of collapse.

Recently, Maeda et al. [13,14] investigated the KSS vector of the
second kind in the tilted case. They assumed the perfect fluid
spacetime obeying a relativistic polytropic equation of state.
Further, they assumed two kinds of polytropic equation of state
and showed that such spacetimes must be vacuum in both cases. They
studied the case in which a KSS vector is not only tilted to the
fluid flow but also parallel or orthogonal. In the recent paper
[15], the same authors discussed the classification of the
spherically symmetric KSS perfect fluid and dust solutions. This
analysis has provided some interesting solutions. In this paper,
we shall use the same procedure to calculate self-similar
solutions for the cylindrically symmetric spacetimes. The
governing equations for perfect fluid cosmological models are
introduced and a set of integrability conditions for the existence
of a KSS solutions are derived.

The paper can be outlined as follows. In section 2, we shall
discuss KSS of different kinds for the cylindrically symmetric
spacetimes. Section 3 is devoted to titled perfect fluid case. In
section 4, we shall find out the titled dust solutions. Sections 5
and 6 are used to discuss the orthogonal perfect fluid and dust
solutions respectively. Finally, we shall summarise and discuss
all the results.

\section{Cylindrically Symmetric Spacetime and Kinematic Self-Similarity}

The general cylindrically symmetric spacetime is given by the line
element [16]
\begin{equation}
ds^2=-e^{2\phi(t,r)}dt^2+ dr^2+e^{2\mu(t,r)}d\theta^2+
e^{2\nu(t,r)}dz^2,
\end{equation}
where $\phi$, $\mu$ and $\nu$ are arbitrary functions of $t$ and
$r$. The energy-momentum tensor for a perfect fluid is given by
\begin{equation}
T_{ab}=[\rho(t,r)+p(t,r)]u_{a}u_{b}+ p(t,r)g_{ab},\quad
(a,b=0,1,2,3).
\end{equation}
where $\rho$ is the density, $p$ is the pressure and $u_{a}$ is
the four velocity of the fluid element in the co-moving coordinate
system given as $u_{a}=(-e^{\phi},0,0,0)$. The EFEs reduce to the
following form [17]
\begin{eqnarray}
8\pi G \rho&=& e^{-2\phi}\nu_{t}\mu_{t}-\nu_{r}\mu_{r}-{\mu_{r}}^{2}
-\mu_{rr}-{\nu_{r}}^{2}-\nu_{rr},\\
8\pi Gp
&=&e^{-2\phi}(-\nu_{tt}-\mu_{tt}+\phi_{t}\mu_{t}+\phi_{t}\nu_{t}
-{\mu_{t}}^{2}-\nu_{t}\mu_{t}-{\nu_{t}}^{2})
+\phi_{r}\mu_{r} \nonumber\\ & & +\phi_{r}\nu_{r} +\nu_{r}\mu_{r},\\
8\pi Gp&=& e^{-2\phi}(-\nu_{tt}+\phi_{t}\nu_{t}-{\nu_{t}}^{2})
+{\nu_{r}}^{2}+\nu_{rr}+\phi_{r}\nu_{r}+{\phi_{r}}^{2}+\phi_{rr},\\
8\pi Gp&=& e^{-2\phi}(-\mu_{tt}+\phi_{t}\mu_{t}-{\mu_{t}}^{2})
+{\mu_{r}}^{2}+\mu_{rr}+\phi_{r}\mu_{r}+{\phi_{r}}^{2}+\phi_{rr},\\
0&=&\mu_{tr}+\mu_{t}\mu_{r}-\phi_{r}\mu_{t}+\nu_{tr}+\nu_{t}\nu_{r}
-\phi_{r}\nu_{t}.
\end{eqnarray}
The conservation of energy-momentum tensor ${T^{ab}}_{;b}=0$
yields the following equations
\begin{equation}
\mu_{t}=-\frac{\rho_{t}}{(\rho+p)}-\nu_{t},
\end{equation}
and
\begin{equation}
\phi_{r}=-\frac{\rho_{r}}{(\rho+p)}.
\end{equation}
For a cylindrically symmetric spacetime, the general form of a
vector field $\eta$ can be written as
\begin{equation}
\eta^{a}\frac{\partial}{\partial
x^a}=h_{1}(t,r)\frac{\partial}{\partial
t}+h_{2}(t,r)\frac{\partial}{\partial r},
\end{equation}
where $h_{1}$ and $h_{2}$ are arbitrary functions. When $\eta$ is
parallel to the fluid flow, $h_{2}=0$ while $h_{1}=0$ indicates
that $\eta$ is orthogonal to the fluid flow. When $\eta$ is tilted
to the fluid flow, both $h_{1}$ and $h_{2}$ are non-zero.

A KSS vector $\xi$ satisfies the following two conditions
\begin{eqnarray}
\pounds_{\xi}h_{ab}&=& 2\delta h_{ab},\\
\pounds_{\xi}u_{a}&=& \alpha u_{a},
\end{eqnarray}
where $h_{ab}=g_{ab}+u_au_b$ is the projection tensor, $\alpha$
and $\delta$ are constants. The similarity transformation is
characterized by the scale independent ratio, $\alpha/\delta$,
which is known as the similarity index. The similarity index gives
rise to the following two cases according as:
\begin{eqnarray*}
1. \quad \delta\neq0,\\
2. \quad \delta=0.
\end{eqnarray*}
{\bf Case 1:} If $\delta\neq0$ it can be chosen as unity and the
KSS vector for the titled case can take the following form
\begin{equation}
\xi^{a}\frac{\partial}{\partial x^a}=(\alpha
t+\beta)\frac{\partial}{\partial t}+r\frac{\partial}{\partial r}.
\end{equation}
For this case, the similarity index, $\alpha/\delta$, further
yields the following three different possibilities.
\begin{eqnarray*}
(i)\quad\delta\neq0,\quad\alpha=1\quad
(\beta~can~be~taken~to~be~zero),\\
(ii)\quad\delta\neq0,\quad\alpha=0\quad
(\beta~can~be~taken~to~be~unity),\\
(iii)\quad\delta\neq0,\quad\alpha\neq0,1\quad
(\beta~can~be~taken~to~be~zero).
\end{eqnarray*}
The case 1(i) corresponds to the self-similarity of the {\it first
kind}. In this case $\xi$ is a homothetic vector and the
self-similar variable $\xi$ turns out to be $r/t$. For the second
case 1(ii), it is termed as the self-similarity of the {\it zeroth
kind} and the self-similar variable follows
\begin{eqnarray*}
\xi=r e^{-t}.
\end{eqnarray*}
In the last case 1(iii), it is called the self-similarity of the
{\it second kind} and the self-similar variable becomes
\begin{eqnarray*}
\xi=\frac{r}{(\alpha t)^\frac{1}{\alpha}}.
\end{eqnarray*}
It turns out that for the case (1), when $\delta\neq0$, with the
self-similar variable $\xi$, the metric functions become
\begin{equation}
\phi=\phi(\xi),\quad e^{\mu}=re^{\mu(\xi)},\quad
e^{\nu}=re^{\nu(\xi)}.
\end{equation}
The case (2), in which $\delta=0$ and $\alpha\neq0$ ($\alpha$ can
be unity and $\beta$ can be re-scaled to zero), the
self-similarity is known as {\it infinite kind}. In this case, the
KSS vector $\xi$ turns out to be
\begin{equation}
\xi^{a}\frac{\partial}{\partial x^a}=t\frac{\partial}{\partial
t}+r\frac{\partial}{\partial r}
\end{equation}
and the self-similar variable will become
\begin{eqnarray*}
\xi=e^{\frac{r}{c}}/t,
\end{eqnarray*}
where $c$ is an arbitrary constant. The metric functions will take
the following form
\begin{equation}
\phi=\phi(\xi),\quad\mu=\mu(\xi),\quad\nu=\nu(\xi).
\end{equation}
It is mentioned here that for the cylindrically symmetric
spacetime, the self-similar variable of the first, second and
zeroth kinds are the same as for the spherically symmetric
spacetime but for the infinite kind we obtain different
self-similar variable. Further, it is noted that for
$\delta=0=\alpha$, the KSS vector $\xi$ becomes Killing vector.

If the KSS vector $\xi$ is parallel to the fluid flow, it follows
that
\begin{equation}
\xi^{a}\frac{\partial}{\partial x^a}= f(t)\frac{\partial}{\partial
t},
\end{equation}
where $f(t)$ is an arbitrary function and the self-similar
variable is $r$. In this case, we obtain contradictory results for
the cylindrically symmetric metric in all kinds. This implies that
there does not exist any solution for $\xi$ parallel to the fluid
flow. If the KSS vector $\xi$ is orthogonal to the fluid flow, it
takes the following form
\begin{equation}
\xi^{a}\frac{\partial}{\partial x^a}= g(r)\frac{\partial}{\partial
r},
\end{equation}
where $g$ is an arbitrary function and the self-similar variable
is $t$.

We assume the following two types of polytropic equation of states
(EOS). We denote the first equation of state by EOS(1) and is
given by
\begin{eqnarray*}
p=k\rho^{\gamma},
\end{eqnarray*}
where $k$ and $\gamma$ are constants. The other EOS can be written
as [12]
\begin{eqnarray*}
p=kn^{\gamma},\\
\rho=m_{b}n+\frac{p}{\gamma-1},
\end{eqnarray*}
where $m_{b}$ is a constant and corresponds to the baryon mass,
and $n(t,r)$ corresponds to baryon number density. This equation
is called second equation of state written as EOS(2). For EOS(1)
and EOS(2), we take $k\neq0$ and $\gamma\neq0,1$. The third
equation of state, denoted by EOS(3), is the following
\begin{eqnarray*}
p=k\rho.
\end{eqnarray*}
Here we assume that $-1\leq k \leq 1$ and $k \neq 0.$

For different values of $\gamma$, EOS(1) and EOS(2) have different
properties. Thermodynamical instability of the fluid is shown for
$\gamma<0.$ For $0<\gamma<1,$ both EOS(1) and EOS(2) are
approximated by a dust fluid in high density regime. For
$\gamma>1$, EOS(2) is approximated by EOS(3) with $k= \gamma-1$ in
high density regime. The cases $\gamma>2$ for EOS(2) and
$\gamma>1$ for EOS(2) shows that the dominant energy condition can
be violated in high density regime which is physically not
interesting [14].

\section{Tilted Case for Perfect Fluid}

\subsection{Self-similarity of the second kind}

Here we discuss the self-similarity of the second kind for the
tilted perfect fluid case. In this case, it follows from the EFEs
that the energy density $\rho$ and pressure $p$ must take the
following form
\begin{eqnarray}
8\pi G\rho &=& \frac{1}{r^2}[\rho_1(\xi)+\frac{r^2}{t^2}\rho_2(\xi)],\\
8\pi Gp &=& \frac{1}{r^2}[p_1(\xi)+\frac{r^2}{t^2}p_2(\xi)],
\end{eqnarray}
where the self-similar variable is $\xi=r/(\alpha
t)^\frac{1}{\alpha}$. If the EFEs and the equations of motion for
the matter field are satisfied for $O[(\frac{r}{t})^0]$ and
$O[(\frac{r}{t})^2]$ terms separately, we obtain a set of ordinary
differential equations. Thus Eqs.(6)-(12) reduce to the following
\begin{eqnarray}
-\rho_1 &=&
1+\dot{\mu}\dot{\nu}+\ddot{\mu}+{\dot{\mu}}^2+2\dot{\mu}
+\ddot{\nu}+{\dot{\nu}}^2+2\dot{\nu},\\
\alpha^2 \rho_2 &=& e^{-2\phi}\dot{\mu}\dot{\nu},\\
p_1 &=& 1+2\dot{\phi}+\dot{\phi}\dot{\nu}+\dot{\phi}\dot{\mu}
+\dot{\nu}+\dot{\mu}+\dot{\mu}\dot{\nu},\\
-\alpha^2e^{2\phi}p_2&=&\ddot{\nu}+{\dot{\nu}}^2+\alpha\dot{\nu}
+\ddot{\mu}+{\dot{\mu}}^2+\dot{\mu}-\dot{\phi}\dot{\nu}
-\dot{\phi}\dot{\mu}+\dot{\mu}\dot{\nu},\\
0&=&\ddot{\nu}+{\dot{\nu}}^2+\dot{\nu}+\ddot{\mu}+{\dot{\mu}}^2
+\dot{\mu}-\dot{\phi}\dot{\nu}-\dot{\phi}\dot{\mu},\\
\dot{\phi}(\rho_1+p_1) &=& 2p_1-\dot{p_1},\\
\dot{\phi}(\rho_2+p_2) &=& -\dot{p_2},\\
-(\dot{\mu}+\dot{\nu})(\rho_1+p_1) &=& \dot{\rho_1},\\
-(\dot{\mu}+\dot{\nu})(\rho_2+p_2) &=& \dot{\rho_2}+2\alpha
\rho_2,
\end{eqnarray}
where dot $(.)$ represents derivative with respect to $ln(\xi).$
Adding Eqs.(24) and (26) and using Eq.(28), we get
\begin{equation}
\rho_1+p_1= 2\dot{\phi},
\end{equation}
Using Eq.(33) in Eq.(29) and Eq.(30), we have
\begin{equation}
(\rho_1+p_1)^2= 4p_1-\dot{p_1},
\end{equation}
and
\begin{equation}
(\rho_2+p_2)(\rho_1+p_1)= -2\dot{p_2}
\end{equation}
respectively.

\subsubsection{Equations of State (1) and (2)}

If a perfect fluid satisfies EOS(1) for $k\neq0$ and
$\gamma\neq0,1$, Eqs.(22) and (23) imply that
\begin{equation}
\alpha=\gamma,\quad p_1=\rho_2=0,\quad p_2= \frac{k}{(8\pi
G)^{(\gamma-1)}\gamma^2}\xi^{-2\gamma}{\rho_1}^\gamma,
\quad[Case~I]
\end{equation}
or
\begin{equation}
\alpha=\frac{1}{\gamma},\quad p_2=\rho_1=0,\quad
p_1=\frac{k}{(8\pi
G)^{(\gamma-1)}\gamma^{2\gamma}}\xi^{2}{\rho_2}^\gamma.\quad[Case~II]
\end{equation}
If a perfect fluid obeys EOS(2) for $k\neq0$ and $\gamma\neq0,1$,
we find from Eqs.(22) and (23) that
\begin{equation}
\alpha=\gamma,\quad p_1=0,\quad p_2=\frac{k}{{m_b}^{\gamma}(8\pi
G)^{(\gamma-1)}\gamma^2}\xi^{-2\gamma}{\rho_1}^\gamma=(\gamma-1)\rho_2,
\quad [Case~III]
\end{equation}
or
\begin{equation}
\alpha=\frac{1}{\gamma}, \quad p_2=0, \quad
p_1=\frac{k}{{m_b}^{\gamma}(8\pi
G)^{(\gamma-1)}\gamma^{2\gamma}}\xi^{2}{\rho_2}^\gamma=(\gamma-1)\rho_1.
\quad [Case~IV]
\end{equation}
In the cases I and III, Eq.(34) gives $\rho_1=0$ and ultimately we
have a vacuum spacetime. For the cases II and IV, using Eq.(35),
it can also be shown that the spacetime is vacuum. Thus we can
conclude that the spacetime must be vacuum in all these cases.

\subsubsection{Equation of State (3)}

When a perfect fluid satisfies EOS(3), it follows from Eqs.(22)
and (23) that
\begin{equation}
p_1=k\rho_1, \quad p_2=k\rho_2. \quad [Case~V]
\end{equation}
For $k=-1$, we use Eqs.(29), (30) and (32) which ultimately yields
a vacuum spacetime. When $k\neq-1$, we assume that $\rho_1\neq0$
and $\rho_2\neq0$. In this case, using Eqs.(29) and (30), it
follows that
$2-\frac{\dot{\rho_1}}{\rho_1}=-\frac{\dot{\rho_2}}{\rho_2},$ and
from Eqs.(31) and (32),  we obtain
$2\alpha+\frac{\dot{\rho_2}}{\rho_2}=\frac{\dot{\rho_1}}
{\rho_1}.$ These two expressions imply that $\rho_1\rho_2=0$ as
$\alpha\neq1$. For the case when $\rho_1=0=p_1$ and $\rho_2\neq0$,
we have a contradiction.

In the case when $\rho_2=0=p_2$ and $\rho_1\neq0$, we subtract
Eq.(28) from Eq.(27) and using Eq.(25) so that
\begin{equation}
(\alpha-1)(\dot{\mu}+\dot{\nu})= 0.
\end{equation}
This implies that $\dot{\mu}=-\dot{\nu}$ as $\alpha\neq0$. Using
this in Eq.(31) and making use of EOS(III), we have $\dot{\rho}=0$
which means $\rho_1=conatant=w_0$ and this gives
$p_1=p_0=constant$. Using these results in Eq.(29), we have
\begin{equation}
\dot{\phi}= \frac{2k}{k+1},
\end{equation}
and this gives
\begin{equation}
e^\phi=c_0\xi^{\frac{2k}{k+1}}.
\end{equation}
From Eqs.(41) and Eq.(28) we find $\mu$ to be a constant and since
$\dot{\mu}=-\dot{\nu}$, therefore we can say $\nu=constant$. The
resulting solution becomes
\begin{eqnarray}
e^\phi=c_0\xi^{\frac{2k}{k+1}},\quad e^\mu=a_0, \quad e^\nu=b_0,
\nonumber\\
w_0=-1,\quad p_1=-k, \quad k=-3\pm 2\sqrt{2}.
\end{eqnarray}
where $a_0$, $b_0$, $c_0$ are constants.

\subsection{Self-similarity of the zeroth kind}

In this section we shall attempt self-similar solutions of the
zeroth kind. For this case, the EFEs indicate that the quantities
$\rho$ and $p$ should be of the form
\begin{eqnarray}
8 \pi G \rho &=& \frac{1}{r^2}[\rho_1(\xi)+r^2\rho_2(\xi)],\\
8 \pi Gp &=& \frac{1}{r^2}[p_1(\xi)+r^2p_2(\xi)],
\end{eqnarray}
where the self-similar variable is $\xi=\frac{r}{e^{-t}}$. If it
is assumed that the EFEs and the equations of motion for the
matter field are satisfied for $O[(r)^0]$ and $O[(r)^2]$ terms
separately, we obtain the following set of ordinary differential
equations.
\begin{eqnarray}
-\rho_1&=&1+\dot{\mu}\dot{\nu}+\ddot{\mu}+{\dot{\mu}}^2+2\dot{\mu}
+\ddot{\nu}+{\dot{\nu}}^2+2\dot{\nu},\\
\rho_2 &=& e^{-2\phi}\dot{\mu}\dot{\nu},\\
p_1 &=& 1+2\dot{\phi}+\dot{\phi}\dot{\nu}+\dot{\phi}\dot{\mu}
+\dot{\nu}+\dot{\mu}+\dot{\mu}\dot{\nu},\\
e^{2\phi}p_2&=&\ddot{\nu}+{\dot{\nu}}^2+\ddot{\mu}+{\dot{\mu}}^2
-\dot{\phi}\dot{\nu}-\dot{\phi}\dot{\mu}+\dot{\mu}\dot{\nu},\\
0&=&\ddot{\nu}+{\dot{\nu}}^2+\dot{\nu}+\ddot{\mu}+{\dot{\mu}}^2
+\dot{\mu}-\dot{\phi}\dot{\nu}-\dot{\phi}\dot{\mu},\\
\dot{\phi}(\rho_1+p_1) &=& 2p_1-\dot{p_1},\\
\dot{\phi}(\rho_2+p_2) &=& -\dot{p_2},\\
(\dot{\mu}+\dot{\nu})(\rho_1+p_1) &=& \dot{\rho_1},\\
(\dot{\mu}+\dot{\nu})(\rho_2+p_2) &=& \dot{\rho_2},\\
p_1 &=& \ddot{\phi}+{\dot{\phi}}^2+\dot{\phi}\dot{\nu}
+\ddot{\nu}+{\dot{\nu}}^2+\dot{\nu},\\
e^{2\phi}p_2&=&\dot{\phi}\dot{\nu}-\ddot{\nu}-{\dot{\nu}}^2.
\end{eqnarray}
Here again dot $(.)$ represents derivative with respect to
$ln(\xi)$. If we add Eqs.(47) and (49) and use Eq.(51), we get
\begin{equation}
\rho_1+p_1= 2\dot{\phi}.
\end{equation}
Substituting Eq.(58) in Eqs.(52) and (53), it follows that
\begin{equation}
(\rho_1+p_1)^2= 4p_1-2\dot{p_1},
\end{equation}
and
\begin{equation}
(\rho_2+p_2)(\rho_1+p_1)= -2\dot{p_2}
\end{equation}
respectively.

\subsubsection{EOS(1) and EOS(2)}

For these two EOS, we obtain a contradiction and consequently we
do not have any solution.

\subsubsection{EOS(3)}

When a perfect fluid satisfies EOS(3), it follows from Eqs.(45)
and (46) that
\begin{equation}
p_1=k\rho_1,\quad  p_2=k\rho_2.\quad  [Case~III]
\end{equation}
First we assume that $\rho_1\neq0$ and $\rho_2\neq0$. For this
case, we make use of Eqs.(52) and (53) so that we get
$2\rho_1\rho_2-\dot{\rho_1}{\rho_2}+\dot{\rho_2}{\rho_1}=0$. Also
from Eqs.(55) and (56), we obtain
$-\dot{\rho_1}{\rho_2}+\dot{\rho_2}{\rho_1}=0$. Subtracting these
two expressions we can conclude that $\rho_1\rho_2=0$ which gives
a contradiction. If we assume that $\rho_1=0=p_1,~\rho_2\neq0$, we
again have a contradiction.

In the third case, we assume that $\rho_2=0=p_2,~\rho_1\neq0$. It
follows from Eq.(48) that $\dot{\mu}\dot{\nu}=0$ and also from
Eqs.(50) and (51) we get $\dot{\mu}+\dot{\nu}=0$. Eq.(54) requires
that $\rho_1=w_0=constant$ which implies that $p_1=k w_0$ and
ultimately Eq.(49) gives the value of $\phi.$ Here we require that
$k\neq-1.$ The resulting solution is
\begin{eqnarray}
e^{\phi}=c_0 \xi^{\frac{-(1+k)}{2}},\quad e^\mu=1=e^\nu,\nonumber\\
w_0=-1,\quad p_0=-k,\quad k=-3\pm 2\sqrt{2}.
\end{eqnarray}
This corresponds to the solution already found in the second kind
with EOS(3).

\subsection{Self-similarity of the infinite kind}

This section is devoted to discuss the self-similar solution of
the infinite kind. In this case, the EFEs imply that the
quantities $\rho$ and $p$ must be of the form
\begin{eqnarray}
8 \pi G \rho &=& \frac{1}{t^2}\rho_1(\xi)+\rho_2(\xi),\\
8 \pi Gp &=& \frac{1}{t^2}p_1(\xi)+p_2(\xi),
\end{eqnarray}
where $\xi=\frac{e^\frac{r}{c}}{t}.$  Now if we require that the
EFEs and the equations of motion for the matter field are
satisfied for $O[(t)^0]$ and $O[(t)^{-2}]$ terms separately, we
obtain a set of ordinary differential equations. For a perfect
fluid, Eqs.(6)-(12) takes the following form
\begin{eqnarray}
\rho_1 &=& e^{-2\phi}\dot{\mu}\dot{\nu},\\
-c^2\rho_2 &=& \dot{\mu}\dot{\nu}+\ddot{\mu}+{\dot{\mu}}^2
+\ddot{\nu}+{\dot{\nu}}^2,\\
-e^{2\phi}p_1&=&\ddot{\nu}+{\dot{\nu}}^2+\dot{\nu}+\ddot{\mu}
+{\dot{\mu}}^2+\dot{\mu}-\dot{\phi}\dot{\nu}
-\dot{\phi}\dot{\mu}+\dot{\mu}\dot{\nu},\\
c^2p_2 &=& \dot{\phi}\dot{\nu}+\dot{\phi}\dot{\mu}
+\dot{\mu}\dot{\nu},\\
0&=&\ddot{\nu}+{\dot{\nu}}^2+\ddot{\mu}+{\dot{\mu}}^2
-\dot{\phi}\dot{\nu}-\dot{\phi}\dot{\mu},\\
\dot{\phi}(\rho_1+p_1) &=& -\dot{p_1},\\
\dot{\phi}(\rho_2+p_2) &=& -\dot{p_2},\\
(\dot{\mu}+\dot{\nu})(\rho_1+p_1) &=& -\dot{\rho_1},\\
(\dot{\mu}+\dot{\nu})(\rho_2+p_2) &=& \dot{\rho_2},\\
-e^{2\phi}p_1&=&\ddot{\nu}+{\dot{\nu}}^2+\dot{\nu}-\dot{\phi}\dot{\nu},\\
c^2p_2&=&
\ddot{\phi}+{\dot{\phi}}^2+\dot{\phi}\dot{\nu}+\ddot{\nu}+{\dot{\nu}}^2,\\
-e^{2\phi}p_1&=&\ddot{\mu}+{\dot{\mu}}^2+\dot{\mu}-\dot{\phi}\dot{\mu},\\
c^2p_2&=&
\ddot{\phi}+{\dot{\phi}}^2+\dot{\phi}\dot{\mu}+\ddot{\mu}+{\dot{\mu}}^2,
\end{eqnarray}
where dot $(.)$ represents derivative with respect to $ln(\xi)$.
Now if we subtract Eq.(67) from Eq.(65) and use Eq.(69), and also
subtract Eq.(66) from Eq.(68) and use Eq.(69), it follows that
\begin{eqnarray}
-e^{2\phi}(\rho_1+p_1)=\dot{\mu}+\dot{\nu},\\
\rho_2+p_2= 0
\end{eqnarray}
respectively. From Eq.(71) we can write $p_2=p_0=constant,$ and
Eq.(73) gives $\rho_2=w_0=constant.$ This implies that $p_0=-w_0.$

\subsubsection{EOS(1) and EOS(2)}

When a perfect fluid satisfies EOS(1), it can be seen from Eq.(63)
and Eq.(64) that
\begin{eqnarray}
p_1=\rho_1=0,\nonumber\\
p_2= k(8\pi G)^{(1-\gamma)}{\rho_2}^\gamma.\quad [Case~I]
\end{eqnarray}
For the condition given by EOS(2), it turns out that
\begin{eqnarray}
p_1=0=\rho_1,\nonumber\\ p_2=\frac{k}{{m_b}^{\gamma}(8\pi
G)^{(\gamma-1)}}{(\rho_2-\frac{p_2}{( \gamma-1)})}^\gamma.\quad
[Case~II]
\end{eqnarray}
In both cases, we have $w_0=0=p_0$ and consequently spacetime
turns out to be vacuum.

\subsubsection{EOS(3)}

For this equation of state, it follows from Eqs.(63) and (64) that
\begin{equation}
p_1=k\rho_1,\quad  p_2=k\rho_2.\quad [Case~III]
\end{equation}
Eq.(77) implies that $p_2=-\rho_2,$ and from Eq.(71) with Eq.(73),
we have $p_2=p_0,~\rho_0=w_0$ which yield $p_0=-w_0$. This gives
rise to the following two cases either $k=-1$ or $p_0=0=w_0$. In
the first case, we have
\begin{equation}
p_1+\rho_1=0.
\end{equation}
If we make use of Eqs.(70) and (72) with Eq.(83), it follows that
$\rho_1=constant$ and$p_1=constant$. Eqs.(72) and (66) together
give $\dot{\mu}=-\dot{\nu}$ and on using this in Eq.(69), it
follows that $\dot{\mu}=0=\dot{\nu}$. The resulting solution is
\begin{eqnarray}
e^{\phi}=ln(\xi),\quad e^\mu=1, \quad e^\nu=1,\nonumber\\
\rho_1=\rho_2=0,\quad p_1=p_2=0,\quad k=-1.
\end{eqnarray}
This corresponds to Minkowski spacetime.

For the second case, if we use Eqs.(74) and (76) in Eq.(67), we
get
\begin{equation}
-e^{2\phi}p_1=\dot{\mu}\dot{\nu}.
\end{equation}
Also, use of Eqs.(74) and (76) in Eq.(69) yield
\begin{equation}
-e^{2\phi}p_1=\dot{\mu}+\dot{\nu}.
\end{equation}
Using these two in Eq.(68), we further have two possibilities
either $p_1=0$ or $\dot{\phi}=\frac{1}{2}$. For the first
possibility we have the same spacetime as given above but for the
second possibility we get a contradiction.

\section{Tilted Dust Case}

\subsection{Self-similarity of the second kind}

If we set $p_1=0=p_2$ in the basic Eqs.(14)-(22) for the tilted
perfect fluid case (self-similarity of the second kind), Eqs.(19)
and (20) immediately gives $\dot{\phi}=0$ and we can take
$e^\phi=c_0$. The rest of the equations reduce to
\begin{eqnarray}
-\rho_1 &=&
1+\dot{\mu}\dot{\nu}+\ddot{\mu}+{\dot{\mu}}^2+2\dot{\mu}
+\ddot{\nu}+{\dot{\nu}}^2+2\dot{\nu},\\
\alpha^2 \rho_2{c_0}^2 &=& \dot{\mu}\dot{\nu},\\
0 &=& 1+\dot{\nu}+\dot{\mu}+\dot{\mu}\dot{\nu},\\
-\alpha^2{c_0}^2p_2&=&\ddot{\nu}+{\dot{\nu}}^2
+\alpha\dot{\nu}+\ddot{\mu}+{\dot{\mu}}^2+\dot{\mu}
+\dot{\mu}\dot{\nu},\\
0&=&\ddot{\nu}+{\dot{\nu}}^2+\dot{\nu}+\ddot{\mu}
+{\dot{\mu}}^2+\dot{\mu},\\
-(\dot{\mu}+\dot{\nu})\rho_1 &=& \dot{\rho_1},\\
-(\dot{\mu}+\dot{\nu})\rho_2 &=& \dot{\rho_2}+2\alpha \rho_2,\\
0&=&\ddot{\nu}+2\dot{\nu},\\
0&=&\ddot{\nu}+{\dot{\nu}}^2+\alpha \dot{\nu},\\
0&=&\ddot{\mu}+2\dot{\mu},\\
0&=&\ddot{\mu}+{\dot{\mu}}^2+\alpha \dot{\mu}.
\end{eqnarray}
Eqs.(91) and (89) with Eq.(87) gives $\rho_1=0.$ Now making use of
Eqs.(94) and (97) in Eq.(90), we have $\rho_2=0,~
\dot{\mu}\dot{\nu}=0,~\rho_2=0$. This leads to contradiction to
our assumption that $\alpha\neq 1$.

\subsection{Self-similarity of zeroth kind}

If we choose $p_1=0=p_2$ in Eqs.(51)-(57) for the tilted perfect
fluid case (self-similarity of zeroth kind), it follows from
Eqs.(52) and (53) that $ \dot{\phi}=0$ and we can take
$e^\phi=c_0$. Now Eqs.(56) and (57) show that $\nu$ is a constant
but then Eqs.(49) and (50) give contradiction. Thus we can
conclude that there is no solution in this case.

\subsection{Self-similarity of infinite kind}

When we set $p_1=0=p_2$ in Eqs.(65)-(77) for the tilted perfect
fluid case (self-similarity of infinite kind), Eqs.(70) and (71)
imply that $ \dot{\phi}=0$ and we set $e^\phi=c_0$. Now Eqs.(65)
and (68) show that $\rho_1=0,$ and also the two equations Eqs.(69)
and (66) with Eq.(68) give $\rho_2=0$.  This case yields vacuum
spacetime.

\section{Orthogonal Case for Perfect Fluid}

\subsection{Self-similarity of the second kind}

Here we discuss self-similar solution for the orthogonal perfect
fluid case. First, we consider the self-similarity of the second
kind. For this case, the self-similar variable can be written as
\begin{equation}
\xi^{a}\frac{\partial}{\partial x^a}=r\frac{\partial}{\partial r}
\end{equation}
The cylindrically symmetric spacetime takes the form
\begin{equation}
ds^2=-r^{2\alpha}dt^2+ dr^2+r^2e^{2\mu(t)}d\theta^2+r^2
e^{2\nu(t)}dz^2.
\end{equation}
EFEs imply that the quantities $\rho$ and $p$ must be of the form
\begin{eqnarray}
8 \pi G \rho &=& r^{-2}\rho_1(\xi)+r^{-2\alpha}\rho_2(\xi),\\
8 \pi Gp &=& r^{-2}p_1(\xi)+r^{-2\alpha}p_2(\xi),
\end{eqnarray}
where the self-similar variable is $\xi=t$. We note that the
solution is always singular at $r=0$ which corresponds to the
physical center. When the EFEs and the equations of motion for the
matter field are satisfied for $O[(r)^0]$ and $O[(r)^{2-2\alpha}]$
terms separately, we obtain a set of ordinary differential
equations. These are given as
\begin{eqnarray}
\rho_1 &=&-1,\\
\rho_2 &=& e^{-2\phi}\mu'\nu',\\
p_1 &=& 1+2\alpha,\\
e^{2\phi}p_2&=&-\nu''-{\nu'}^2-\mu''-{\mu'}^2+\phi'\nu'+\phi'\mu'
-\mu'\nu',\\
0&=&(1-\alpha)(\nu'+\mu'),\\
(2-\alpha)p_1 &=& \alpha \rho_1,\\
\rho_2 &=& p_2,\\
(\mu'+\nu')[\rho_1+p_1] &=&0,\\
(\mu'+\nu')[\rho_2+p_2] &=& - {\rho_2}'.
\end{eqnarray}
Here prime ($'$) denotes derivative with respect to $t$. since
$\rho_1=0$ contradicts Eq.(102), a vacuum spacetime is not
compatible with this case.

\subsubsection{EOS(1) and EOS(2)}

For a perfect fluid satisfying EOS(I), it follows from Eqs.(100)
and (101) that
\begin{eqnarray}
\alpha=\gamma,\quad p_1=\rho_2=0, \nonumber\\
p_2= \frac{k}{(8\pi G)^{(\gamma-1)}}{\rho_1}^\gamma, \quad
[Case~I]
\end{eqnarray}
or
\begin{eqnarray}
\alpha=\frac{1}{\gamma},\quad p_2=\rho_1=0, \nonumber\\
p_1=\frac{k}{(8\pi G)^{(\gamma-1)}}{\rho_2}^\gamma,\quad [Case~II]
\end{eqnarray}
For EOS(2), Eqs.(100) and (101) imply that
\begin{eqnarray}
\alpha=\gamma,\quad p_1=0, \nonumber\\
p_2= \frac{k}{{m_b}^{\gamma}(8\pi
G)^{(\gamma-1)}}{\rho_1}^\gamma=(\gamma-1)\rho_2, \quad [Case~III]
\end{eqnarray}
or
\begin{eqnarray}
\alpha=\frac{1}{\gamma},\quad p_2=0,\nonumber\\
p_1= \frac{k}{{m_b}^{\gamma}(8\pi
G)^{(\gamma-1)}}{\rho_2}^\gamma=(\gamma-1)\rho_1. \quad [Case~IV]
\end{eqnarray}
Case II directly gives contradiction to Eq.(102). Also, when we
make use of Eq.(108) with Eq.(114), case IV gives contradiction to
Eq.(102). In case I, again using Eqs.(108) and (111), we have a
contradiction. For case III, we have contradiction to Eq.(107) and
hence no solution.

\subsubsection{EOS(3)}

In this case Eqs.(100) and (101) yield that
\begin{equation}
p_1=k\rho_1, \quad p_2=k\rho_2.\quad  [Case~V]
\end{equation}
Now if we use Eqs.(102) and Eq.(108) in Eq.(115), we obtain
$p_1=-1$. Using this value in Eq.(104), it follows that $\alpha=0$
which contradicts our assumption that $\alpha=1$. Thus there is no
self-similar solution of the second kind for the orthogonal
perfect fluid case.

\subsection{Self-similarity of the zeroth kind}

In the case of self-similarity of the zeroth kind, the basic
equations for perfect fluid gives us a contradiction and hence we
have no solution in this case.

\subsection{Self-similarity of the infinite kind}

For the self-similarity of the infinite kind, EFEs imply that the
quantities $\rho$ and $p$ must be of the form
\begin{eqnarray}
8 \pi G \rho &=& e^{-2r}\rho_1(\xi)+\rho_2(\xi),\\
8 \pi Gp &=& e^{-2r}p_1(\xi)+p_2(\xi),
\end{eqnarray}
where $\xi=t.$  A set of ordinary differential equations is
obtained if EFEs and the equations of motion for the matter field
are satisfied for $O[(r)^0]$ and $O[(r)^{-2}]$ terms separately.
In this case, Eqs.(2.3)-(2.9) take the following form
\begin{eqnarray}
\rho_1 &=& e^{-2\phi}\mu'\nu',\\
\rho_2 &=& 0,\\
-e^{2\phi}p_1&=&\nu''+{\nu'}^2+\mu''+{\mu'}^2-\phi'\nu'
-\phi'\mu'+\mu'\nu',\\
p_2 &=& 0,\\
0&=&\nu'+\mu',\\
p_1 &=& \rho_1,\\
(\mu'+\nu')[\rho_1+p_1] &=& -{\rho_1}',
\end{eqnarray}
where prime $(')$ represents derivative with respect to $t$.
Eqs.(122) and (124) yield that $\rho_1$ is a constant and assume
that $\rho_1=\rho_0$. Also, Eq.(122) implies that $\mu'=-\nu'$.

For EOS(1) and EOS (2), we have a contradiction if we take
$p_2=0=\rho_2$ (as given by the above equations). If we take
$p_1=0=\rho_1$, this gives a vacuum spacetime. If we consider
$p_1=0=\rho_2$ or $p_2=0=\rho_1$, we also get a vacuum spacetime.
EOS (3) gives us imaginary complex results.

\section{Orthogonal Dust case}

For the dust case, we substitute $p_1=0=p_2$ in the basic
equations for the orthogonal perfect fluid case. We obtain
contradiction in all the cases, i.e., second, zeroth and infinte
kinds. Hence there does not exist any self-similar solution for
the orthogonal dust case.

\section{Conclusion}

We have attempted to find out KSS perfect fluid and dust solutions
for the cases when KSS vector is tilted, parallel or orthogonal to
the fluid flow with either EOS(1) or EOS(2) or EOS(3). The
parallel case gives a contradiction and hence there is no
self-similar cylindrical symmetric solution for this case (i.e.
second, zeroth or infinite kind).

For the tilted perfect fluid case (self-similarity of the second
kind), EOS(1) and EOS(2) give only vacuum spacetimes. The EOS(3),
for $k=-1$, yields a contradiction and also the cases
$\rho_1=0=p_1$ and $\rho_2 \neq 0 \neq p_2$ give a contradiction.
We obtain only one solution for EOS(3) when $\rho_2=0=p_2$. For
the tilted perfect fluid case with self-similarity of the zeroth
kind, EOS(1) and EOS(2) give contradiction. However, EOS(3)
provides one solution and the remaining possibilities yield either
a contradiction or a vacuum solution. In the tilted perfect fluid
case (self-similarity of the infinite kind), we obtain that the
spacetime must be vacuum for EOS(1) and EOS(2). EOS(3) for $k=-1$
provides one solution which corresponds to Minkowski spacetime.
The remaining case yields a contradiction.

When we solve the tilted dust case with the self-similarity of the
infinite kind, it follows a vacuum solution while the other kinds
yield contradiction. In the orthogonal perfect fluid case with
self-similarity of the infinite kind, we have vacuum solution for
EOS(1) and EOS(2) while EOS(3) do not give a solution. For zeroth
kind we have a contradiction in basic equations. For the
orthogonal dust case we obtain contradiction in all the cases. The
summary of the results can be given below in the form of tables.
\vspace{0.2cm}

{\bf {\small Table 1.}} {\small Perfect fluid kinematic
self-similar solutions for the EOS(1)}.

\vspace{0.1cm}

\begin{center}
\begin{tabular}{|l|l|}
\hline {\bf Self-similarity} & {\bf Solution}
\\ \hline Second kind (tilted) & Vacuum
\\ \hline Second kind (parallel) & None
\\ \hline Second kind (orthogonal) & None
\\ \hline Zeroth kind (titled) & None
\\ \hline Zeroth kind (parallel) & None
\\ \hline Zeroth kind (orthogonal) & None
\\ \hline Infinite kind (tilted) & Vacuum
\\ \hline Infinite kind (parallel) & None
\\ \hline Infinite kind (orthogonal) & Vacuum
\\ \hline
\end{tabular}
\end{center}
\vspace{0.2cm}

{\bf {\small Table 2.}} {\small Perfect fluid kinematic
self-similar solutions for the EOS(2)}.

\vspace{0.1cm}

\begin{center}
\begin{tabular}{|l|l|}
\hline {\bf Self-similarity} & {\bf Solution}
\\ \hline Second kind (tilted) & Vacuum
\\ \hline Second kind (parallel) & None
\\ \hline Second kind (orthogonal) & None
\\ \hline Zeroth kind (titled) & None
\\ \hline Zeroth kind (parallel) & None
\\ \hline Zeroth kind (orthogonal) & None
\\ \hline Infinite kind (tilted) & Vacuum
\\ \hline Infinite kind (parallel) & None
\\ \hline Infinite kind (orthogonal) & Vacuum
\\ \hline
\end{tabular}
\end{center}
\newpage
{\bf {\small Table 3.}} {\small Perfect fluid kinematic
self-similar solutions for the EOS(3)}.

\vspace{0.1cm}

\begin{center}
\begin{tabular}{|l|l|}
\hline {\bf Self-similarity} & {\bf Solution}
\\ \hline Second kind (tilted) & solution given by Eq.(44)
\\ \hline Second kind (parallel) & None
\\ \hline Second kind (orthogonal) & None
\\ \hline Zeroth kind (tilted) & solution given by Eq.(62)
\\ \hline Zeroth kind (parallel) & None
\\ \hline Zeroth kind (orthogonal) & None
\\ \hline Infinite kind (tilted) & Minkowski
\\ \hline Infinite kind (parallel) & None
\\ \hline Infinite kind (orthogonal) & None
\\ \hline
\end{tabular}
\end{center}
It is to be noted that there is only vacuum solution in the case
of dust fluid for the tilted infinite kind. In the remaining
cases, we do not have solution.
\newpage

\begin{description}
\item  {\bf Acknowledgment}
\end{description}

One of us (SA) would like to acknowledge the enabling role of the
Higher Education Commission Islamabad, Pakistan and appreciate its
financial support through {\it Merit Scholarship Scheme for Ph.D.
Studies in Science and Technology (200 Scholarships)}.

\vspace{2cm}

{\bf \large References}

\begin{description}

\item{[1]} Penston, M.V.: Mon. Not. R. Astr. Soc. {\bf144}(1969)425;\\
Larson, R.B.: Mon. Not. R. Astr. Soc. {\bf145}(1969)271;\\
Shu, F.H.: Astrophys. J. {\bf214}(1977)488;\\
Hunter, C.: Astrophys. J. {\bf218}(1977)834.

\item{[2]} Cahill, M. E. and Taub, A. H.:   Commun. Math. Phys. {\bf21}(1971)1.

\item{[3]} Carter, B. and Henriksen, R. N.:  Annales De Physique {\bf14}(1989)47-53.

\item{[4]} Carter, B. and Henriksen, R. N.:  J. Math. Phys. {\bf32}(1991)2580-2597.

\item{[5]} Carr, B.J.: Phys. Rev. {\bf D62}(2000)044022.

\item{[6]} Carr, B. J. and Coley, A. A.: Phys. Rev. {\bf D62}(2000)044023.

\item{[7]} Coley, A. A.: Class. Quant. Grav. {\bf14}(1997)87-118.

\item{[8]} McIntosh, C.B.G.: Gen. Relat. Gravit. {\bf7}(1975)199.

\item{[9]} Benoit, P. M. and Coley, A. A.: Class. Quant. Grav. {\bf15}(1998)2397-2414.

\item{[10]} Carr, B. J., Coley, A. A., Golaith, M., Nilsson, U. S. and Uggla, C.:
Class. Quant. Grav. {\bf18}(2001).

\item{[11]} Carr, B. J., Coley, A. A., Golaith, M., Nilsson, U. S. and Uggla, C.:
Phys. Rev. {\bf D61}(2000).

\item{[12]} Coley, A. A. and Golaith, M.: Class. Quant. Grav. {\bf17}(1998).

\item{[13]} Maeda, H., Harada, T., Iguchi, H. and Okuyama, N.: Phys. Rev.
{\bf D66}(2002)027501.

\item{[14]} Maeda, H., Harada, T., Iguchi, H. and Okuyama, N.: Prog. Theor.Phys.
{\bf108}(2002)819-851.

\item{[15]} Maeda, H., Harada, T., Iguchi, H. and Okuyama, N.: Prog.Theor. Phys.
{\bf110}(2003)25-63.

\item{[16]} Stephani, H., Kramer, D., Maccallum, M., Hoenselaers, C. and Herlt,E.:
\textit{Exact Solutions of Einstein's Field Equations} (Cambridge
University Press, 2003).

\item{[17]} Sharif, M. and Aziz, Sehar: {\it Kinematic Self-Similar Solutions:
Proceedings of the 11th Regional Conference on Mathematical
Physics and IPM Spring Conference} Tehran-Iran, May, 3-6, 2004.
eds. F. Ardalan, S. Rahvar, N. Sadooghi and F. Shojai (to appear).

\item{[18]} Sharif, M. and Aziz, Sehar: Int. J. Mod. Phys. {\bf D} (to appear,
arXiv: gr-qc/0406029).

\end{description}

\end{document}